\documentclass[journal]{IEEEtran}

\ifCLASSINFOpdf
\else
\fi

\usepackage{graphicx}
\usepackage{amssymb}
\usepackage{slashbox}
\usepackage{amsmath}
\usepackage{amsbsy}

\begin{document}

\title{A Unified Analysis Approach for LMS-based Variable Step-Size Algorithms}

\author{Muhammad~O.~Bin~Saeed,~\IEEEmembership{Member,~IEEE}
\thanks{M. O. Bin Saeed is with the Department of Electrical Engineering, Affiliated Colleges at Hafr Al Batin, King Fahd University of Petroleum \& Minerals, Hafr Al Batin 31991, Saudi Arabia e-mail: (mobs@kfupm.edu.sa).}}

\maketitle

\begin{abstract}
The least-mean-squares (LMS) algorithm is the most popular algorithm in adaptive filtering. Several variable step-size strategies have been suggested to improve the performance of the LMS algorithm. These strategies enhance the performance of the algorithm but a major drawback is the complexity in the theoretical analysis of the resultant algorithms. Researchers use several assumptions to find closed-form analytical solutions. This work presents a unified approach for the analysis of variable step-size LMS algorithms. The approach is then applied to several variable step-size strategies and theoretical and simulation results are compared.
\end{abstract}

\textbf{\emph{Index Terms }}-- Variable step-size, least-mean-square algorithms

\section{Introduction}
\label{Intro}

Many algorithms have been proposed for estimation/system identification but the LMS algorithm has been the most popular \cite{Sayedbook} as it is simple and effective. However, a limiting factor of LMS is that if the step-size of the algorithm is kept high then the algorithm converges quickly but the resultant error floor is high. On the other hand, lowering the step-size is results in improvement in the error performance but the speed of the algorithm becomes slow. In order to overcome this problem, various variable step-size (VSS) strategies have been suggested, which have a high step-size initially for fast convergence but then reduce the step-size with time in order to achieve a low error performance \cite{nagumo}-\cite{Gil2014}. Some algorithms are proposed in literature for specific applications \cite{Egiazarian2003},\cite{Benesty2006},\cite{Akhtar2006},\cite{Filho2009}-\cite{Gil2014}. There are several algorithms that are derived from a constraint on the cost function \cite{nagumo},\cite{gollamudi98},\cite{Pazaitis1999},\cite{wei2001},\cite{Benesty2006}.

In general, all VSS algorithms aim to improve performance at the cost of computational complexity. This trade-off is generally acceptable due to the improvement in performance. However, the additional complexity also results in difficulty in analyzing the algorithm. Authors use several basic assumptions to find closed-form solutions for the analysis of these algorithms. Most of these assumptions are similar. However, each algorithm has to be dealt with separately in order to find the steady-state misadjustment, leading to the steady-state excess-mean-square-error (EMSE). Similarly, the mean-square analysis for each algorithm has to be performed individually.

Based on the similarity of the assumption used by the authors of all these VSS algorithms, this work presents a unified approach for the analysis of VSS LMS algorithms. The aim of this work is to perform a generalized analysis for any VSS strategy that is based on the LMS algorithm. This analysis can be applied to most existing as well as any forthcoming VSS algorithms.

The rest of the paper is divided as follows. Section II presents a working system model and problem statement. Section III details the complete theoretical analysis for VSS LMS algorithms. Simulation results are presented in section IV. Section V concludes this work.

\section{System Model}
\label{Sys_Model}

The unknown system is modeled as an FIR filter in the form of a vector, ${\bf w}_o$, of size ($M \times 1$). The input to the unknown system at any given time $i$ is a ($1 \times M$) complex-valued regressor vector, ${\bf u}(i)$. The observed output of the system is a noise corrupted scalar, $d(i)$. The variables of the system are related by
\begin{equation}\label{sys_eq}
d(i) = {\bf u}(i){\bf w}_o + v(i),
\end{equation}
where $v(i)$ is the complex-valued zero-mean additive noise.

The LMS algorithm iteratively estimates the unknown system with an update equation given by
\begin{equation}\label{lms_eq}
{\bf w}(i+1) = {\bf w} (i) + \mu e(i) {\bf u}^* (i),
\end{equation}
where ${\bf w}(i)$ is the estimate of the unknown system vector at time $i$, $e(i) = d(i) - {\bf u}(i){\bf w}(i)$ is the instantaneous error and $(.)^*$ is the complex conjugate transpose operator. The step-size for the update is defined by the variable $\mu$, which is fixed for the LMS algorithm. In case of a variable step-size algorithm, the step-size is also updated iteratively. The VSS LMS update equations are given by
\begin{eqnarray}
{\bf w}(i+1) &=& {\bf w} (i) + \mu(i) e(i) {\bf u}^* (i), \label{vsslms_eq} \\
\mu (i+1) &=& f\{\mu(i)\}, \label{ss_upd}
\end{eqnarray}
where $f\{.\}$ is a function that defines the update equation for the step-size and is different for every VSS algorithm.

While performing the analysis of the LMS algorithm, the input regressor vector is assumed to be independent of the estimated vector. For the VSS algorithms, it is generally assumed that control parameters are chosen such that the step-size and the input regressor vector are asymptotically independent of each other, resulting in a closed-form steady-state solution that closely matches with the simulation results. For some VSS algorithms, the analytical and simulation results are closely matched during the transition stage as well but this is not always the case. The results are still acceptable for all algorithms as a closed-form solution is obtained.

The main objective of this work is to provide a generalized analysis for VSS algorithms, in lieu with the assumptions mentioned above. The results of this analysis can be applied to VSS algorithms in general as will be shown through specific examples.

\section{Proposed analysis}
\label{Proposed}

The weight-error vector is given by
\begin{equation}\label{err_eq}
{\bf \tilde w}(i) = {\bf w}_o - {\bf w}(i).
\end{equation}
Using (\ref{err_eq}) in (\ref{vsslms_eq}) results in
\begin{eqnarray}
\nonumber{\bf \tilde w}(i+1) &=& [{\rm {\bf I}}_M - \mu(i) {\bf u}^*(i){\bf u}(i)]{\bf \tilde w}(i) \\
&& - \mu(i) {\bf u}^*(i) v(i), \label{vsslms_eq_err}
\end{eqnarray}
where ${\rm {\bf I}}_M$ is an identity matrix of size $M$. Before beginning the analysis, another assumption is made, without loss of generality. The input data is assumed to be circular Gaussian. As a result, the auto correlation matrix of the input regressor vector, given by ${\bf R}_u = {\rm E}[{\bf u}^*(i){\bf u}(i)]$, where ${\rm E}[.]$ is the expectation operator, can be decomposed into its component matrices of eigenvalues and eigenvectors, ${\bf R}_u = {\bf T \Lambda T}^*$, where ${\bf T}$ is the matrix of eigenvectors such that ${\bf T}^*{\bf T} = {\rm {\bf I}}_M$ and ${\bf \Lambda}$ is a diagonal matrix containing the eigenvalues. Using the matrix ${\bf T}$, the following transformations are made
\[
\begin{array}{*{20}c}
{{\bf{\bar w}}(i)  = {\bf{T}}^* {\bf{\tilde w}}(i) } & {{\bf{\bar u}}(i)  = {\bf{u}}(i) {\bf{T}}}  \\
\end{array}
\]
The weight-error update equation thus becomes
\begin{equation}\label{weight_err_eq}
{\bf \bar w}(i+1) = [{\rm {\bf I}}_M - \mu(i) {\bf \bar u}^*(i){\bf \bar u}(i)]{\bf \bar w}(i) - \mu(i) {\bf \bar u}^*(i) v(i).
\end{equation}
\vspace{-1cm}

\subsection{Mean Analysis}
\label{mean_an}

Applying the expectation operator to (\ref{weight_err_eq}) results in
\begin{eqnarray}
\nonumber {\rm E}\left[{\bf \bar w}(i+1)\right] \hspace{-0.5cm} &&= {\rm E}\left[ \left\{{\rm {\bf I}}_M - \mu(i) {\bf \bar u}^*(i){\bf \bar u}(i) \right\}{\bf \bar w}(i)\right. \\
\nonumber && \left. \hspace{0.5cm} - \mu(i) {\bf \bar u}^*(i) v(i) \right] \\
&&= \left\{ {\rm {\bf I}}_M - {\rm E} \left[ \mu(i) {\bf \bar u}^*(i){\bf \bar u}(i) \right] \right\} {\rm E} \left[ {\bf \bar w}(i) \right], \label{weight_err_mean1}
\end{eqnarray}
where the data independence assumption is used to separate ${\rm E}[{\bf w}(i)]$ from the rest of the variables. The second term is 0 as additive noise is independent and zero-mean. Using the assumption that the step-size control parameters are chosen in such a way that the step-size and the input regressor data are asymptotically independent, (\ref{weight_err_mean1}) is further simplified as
\begin{eqnarray}
{\rm E} \left[ {\bf \bar w}(i+1) \right] &=& \left\{ {\rm {\bf I}}_M - {\rm E} \left[ \mu(i) \right] {\bf \Lambda} \right\} {\rm E} \left[ {\bf \bar w}(i) \right], \label{weight_err_mean2}
\end{eqnarray}
where ${\rm E} \left[ {\bf \bar u}^*(i){\bf \bar u}(i) \right] = {\bf \Lambda}$. The sufficient condition for stability is evaluated from (\ref{weight_err_mean2}) and is given by
\begin{equation}\label{mu_limit}
0 < {\rm E}\left[ {\mu \left( i \right)} \right] < \frac{2}{{\beta _{max} } },
\end{equation}
where ${\beta_{max}}$ is the maximum eigenvalue of ${\bf \Lambda}$.

\subsection{Mean-Square Analysis}
\label{mean_sq_an}

Taking the expectation of the squared weighted $l_2$-norm of (\ref{weight_err_eq}) yields
\begin{eqnarray}\label{mean_sq_1}
&&{\rm E}\left[ \left\| {\bf \bar w} (i+1) \right\|^2 _{{\bf \Sigma}} \right] \hspace{0cm} \\
\nonumber &&= {\rm E}\left[ {\bf \bar w}^*(i) {\bf \Sigma'} {\bf \bar w}(i) \right] + {\rm E} \left[ \mu^2 (i) v^2 (i) {\bf \bar u}(i) {\bf \Sigma} {\bf \bar u}^* (i) \right]\\
\nonumber && - {\rm E} \left[ \mu (i) v (i) {\bf \bar u} (i) {\bf \Sigma} \left\{{\rm {\bf I}}_M - \mu (i) {\bf \bar u}^* (i) {\bf \bar u} (i) \right\} {\bf \bar w}(i) \right] \\
\nonumber && - {\rm E} \left[ {\bf \bar w}^*(i) \left\{{\rm {\bf I}}_M - \mu (i) {\bf \bar u}^* (i) {\bf \bar u} (i) \right\} {\bf \Sigma} \mu (i) v (i) {\bf \bar u}^* (i) \right],
\end{eqnarray}
where $\left\|.\right\|$ is the $l_2$-norm operator and ${\bf \Sigma}$ is a weighting matrix. The weighting matrix ${\bf \Sigma'}$ is given by
\begin{eqnarray}\label{sigma_prime}
\nonumber {\bf \Sigma'} &=& \left\{{\rm {\bf I}}_M - \mu (i) {\bf \bar u}^* (i) {\bf \bar u} (i) \right\}^* {\bf \Sigma} \left\{{\rm {\bf I}}_M - \mu (i) {\bf \bar u}^* (i) {\bf \bar u} (i) \right\} \\
\nonumber &=& {\rm {\bf I}}_M - \mu (i) {\bf \bar u}^* (i) {\bf \bar u} (i) {\bf \Sigma} - \mu (i) {\bf \Sigma} {\bf \bar u}^* (i) {\bf \bar u} (i) \\
&&+ \mu^2 (i) {\bf \bar u}^* (i) {\bf \bar u} (i) {\bf \Sigma} {\bf \bar u}^* (i) {\bf \bar u} (i)
\end{eqnarray}
The last term in (\ref{mean_sq_1}) is 0 due to independence of additive noise. Using the data independence assumption, the remaining 2 terms are simplified as
\begin{equation}\label{mean_sq_2}
{\rm E}\left[ \left\| {\bf \bar w} (i+1) \right\|^2 _{{\bf \Sigma}} \right] = {\rm E}\left[ \left\| {\bf \bar w} (i) \right\|^2 _{{\bf \Sigma'}} \right] + \sigma^2_v {\rm E} \left[ \mu^2 (i) \right] {\rm Tr} \left\{ {\bf \Sigma \Lambda} \right\},
\end{equation}
where $\sigma^2_v$ is the additive noise variance, ${\rm Tr} \left\{.\right\}$ is the {\em trace} operator and ${\rm E} \left[ {\bf u}(i) {\bf \Sigma} {\bf u}^T (i) \right] = {\rm Tr} \left\{ {\bf \Sigma \Lambda} \right\}$. Once again invoking the data independence assumption, we write ${\rm E}\left[ \left\| {\bf \bar w} (i) \right\|^2 _{{\bf \Sigma'}} \right] = {\rm E}\left[ \left\| {\bf \bar w} (i) \right\|^2 _{{\rm E} \left[ {\bf \Sigma'} \right]} \right]$. Further, taking ${\rm E} \left[ {\bf \Sigma'} \right] = {\bf \Sigma'}$ and simplifying, (\ref{sigma_prime}) is rewritten as
\begin{eqnarray}\label{sigma_prime1}
\nonumber {\bf \Sigma'} &=& {\rm {\bf I}}_M - 2{\rm E} \left[ \mu (i) \right] {\bf \Lambda \Sigma} + {\rm E}\left[\mu^2(i)\right] {\bf \Lambda} {\rm Tr}\left[{\bf \Sigma \Lambda}\right] \\
&&+  {\rm E}\left[\mu^2(i)\right] {\bf \Lambda \Sigma \Lambda}.
\end{eqnarray}
Using the {\em diag} operator, (\ref{mean_sq_2}) and (\ref{sigma_prime1}) are simplified as
\begin{eqnarray}
{\rm E}\left[ \left\| {\bf \bar w}(i+1) \right\|^2 _{{\boldsymbol \sigma}} \right] = {\rm E}\left[ \left\| {\bf \bar w}(i) \right\|^2 _{{\bf F}(i){\boldsymbol \sigma}} \right] + \sigma^2_v {\rm E}\left[\mu^2(i)\right] {\boldsymbol \lambda}^T {\boldsymbol \sigma} \label{mean_sq_3}
\end{eqnarray}
where ${\boldsymbol \sigma} = {\rm diag} \left\{ {\bf \Sigma} \right\}$, ${\boldsymbol \lambda} = {\rm diag}\{{\bf \Lambda}\}$, the weighting matrix ${\bf \Sigma'}$ is replaced with ${\rm diag} \left\{ {\bf \Sigma'} \right\} = {\boldsymbol \sigma'} = {\bf F}(i){\boldsymbol \sigma}$, where ${\bf F}(i)$ is
\begin{equation}\label{F_recur}
{\bf F}(i) = {\rm {\bf I}}_M - 2{\rm E}\left[\mu(i)\right] {\bf \Lambda} + {\rm E}\left[\mu^2(i)\right] \left[{\bf \Lambda}^2  + {\boldsymbol \lambda \boldsymbol \lambda}^T\right].
\end{equation}

Now, using (\ref{mean_sq_3}) and (\ref{F_recur}), the analysis iterates as
\begin{eqnarray*}
{\rm E} \left[ \left\| {\bf \bar w}(0) \right\|^2_{{\boldsymbol \sigma}} \right] \hspace{-0.6cm} &&= \left\| {\bf w}_o \right\|^2_{{\boldsymbol \sigma}}, \\
{\bf F} (0) \hspace{-0.6cm} &&= {\rm {\bf I}}_M - 2\mu(0) {\bf \Lambda} + \mu^2(0) \left[{\bf \Lambda}^2  + {\boldsymbol \lambda \boldsymbol \lambda}^T\right],
\end{eqnarray*}
where ${\rm E} \left[ \mu (0) \right] = \mu(0)$ and ${\rm E} \left[ \mu^2(0) \right] = \mu^2(0)$ as this is the initial step-size value. The first iterative update is given by
\begin{eqnarray*}
{\rm E} \left[ \left\| {\bf \bar w}(1) \right\|^2_{{\boldsymbol \sigma}} \right] \hspace{-0.6cm} &&= {\rm E} \left[ \left\| {\bf \bar w}(0) \right\|^2_{{\bf F}(0){\boldsymbol \sigma}} \right] + \sigma^2_v \mu^2(0) {\boldsymbol \lambda}^T {\boldsymbol \sigma} \\
&&= \left\| {\bf \bar w}_o \right\|^2_{{\bf F}(0){\boldsymbol \sigma}} + \sigma^2_v \mu^2(0) {\boldsymbol \lambda}^T {\boldsymbol \sigma} \\
{\bf F}(1) \hspace{-0.6cm} &&= {\rm {\bf I}}_M - 2{\rm E}\left[\mu(1)\right] {\bf \Lambda} + {\rm E}\left[\mu^2(1)\right] \left[{\bf \Lambda}^2  + {\boldsymbol \lambda \boldsymbol \lambda}^T\right],
\end{eqnarray*}
where the updates ${\rm E} \left[ \mu(1) \right]$ and ${\rm E} \left[ \mu^2(1) \right]$ are obtained from the particular step-size update equation of the VSS algorithm being used. Similarly, the second iterative update is given by
\begin{eqnarray*}
{\rm E} \left[ \left\| {\bf \bar w}(2) \right\|^2_{{\boldsymbol \sigma}} \right] \hspace{-0.6cm} &&= {\rm E} \left[ \left\| {\bf \bar w}(1) \right\|^2_{{\bf F}(1){\boldsymbol \sigma}} \right] + \sigma^2_v {\rm E} \left[ \mu^2(1) \right] {\boldsymbol \lambda}^T {\boldsymbol \sigma} \\
&&= \left\| {\bf \bar w}_o \right\|^2_{{\bf F}(1){\bf F}(0){\boldsymbol \sigma}} + \sigma^2_v \mu^2(0) {\boldsymbol \lambda}^T {\bf F} (1) {\boldsymbol \sigma} \\
&&+ \sigma^2_v {\rm E} \left[ \mu^2(1) \right] {\boldsymbol \lambda}^T {\boldsymbol \sigma} \\
&&= \left\| {\bf \bar w}_o \right\|^2_{{\bf F}(1){\bf F}(0){\boldsymbol \sigma}} \\
&&+ \sigma^2_v {\boldsymbol \lambda}^T \left\{ \mu^2(0) {\bf F} (1) + {\rm E} \left[ \mu^2(1) \right] {\rm {\bf I}}_M \right\} {\boldsymbol \sigma} \\
{\bf F}(2) \hspace{-0.6cm} &&= {\rm {\bf I}}_M - 2{\rm E}\left[\mu(2)\right] {\bf \Lambda} + {\rm E}\left[\mu^2(2)\right] \left[{\bf \Lambda}^2  + {\boldsymbol \lambda \boldsymbol \lambda}^T\right].
\end{eqnarray*}
Continuing, the third iterative update is given by
\begin{eqnarray*}
{\rm E} \left[ \left\| {\bf \bar w}(3) \right\|^2_{{\boldsymbol \sigma}} \right] \hspace{-0.6cm} &&= {\rm E} \left[ \left\| {\bf \bar w}(2) \right\|^2_{{\bf F}(2){\boldsymbol \sigma}} \right] + \sigma^2_v {\rm E} \left[ \mu^2(2) \right] {\boldsymbol \lambda}^T {\boldsymbol \sigma} \\
&&= \left\| {\bf \bar w}_o \right\|^2_{{\bf F}(2){\bf A}(2) {\boldsymbol \sigma}} + \sigma^2_v {\rm E} \left[ \mu^2(2) \right] {\boldsymbol \lambda}^T {\boldsymbol \sigma} \\
&&+ \sigma^2_v {\boldsymbol \lambda}^T \left\{ {\sum\limits_{k = 0}^1 {{\rm E}\left[ {\mu ^2 \left( k \right)} \right]\prod\limits_{m = 2}^{k + 1} {{\bf{F}}\left( m \right)} } } \right\} {\boldsymbol \sigma} \\
{\bf F}(3) \hspace{-0.6cm} &&= {\rm {\bf I}}_M - 2{\rm E}\left[\mu(3)\right] {\bf \Lambda} + {\rm E}\left[\mu^2(3)\right] \left[{\bf \Lambda}^2  + {\boldsymbol \lambda \boldsymbol \lambda}^T\right],
\end{eqnarray*}
where the weighting matrix ${\bf A}(2) = {\bf F}(1){\bf F}(0)$. The fourth iterative update is then given by
\begin{eqnarray*}
{\rm E} \left[ \left\| {\bf \bar w}(4) \right\|^2_{{\boldsymbol \sigma}} \right] \hspace{-0.6cm} &&= \left\| {\bf \bar w}_o \right\|^2_{{\bf F}(3){\bf A}(3) {\boldsymbol \sigma}} + \sigma^2_v {\rm E} \left[ \mu^2(3) \right] {\boldsymbol \lambda}^T {\boldsymbol \sigma} \\
&&+ \sigma^2_v {\boldsymbol \lambda}^T \left\{ {\sum\limits_{k = 0}^2 {{\rm E}\left[ {\mu ^2 \left( k \right)} \right]\prod\limits_{m = 3}^{k + 1} {{\bf{F}}\left( m \right)} } } \right\} {\boldsymbol \sigma} \\
{\bf F}(4) \hspace{-0.6cm} &&= {\rm {\bf I}}_M - 2{\rm E}\left[\mu(4)\right] {\bf \Lambda} + {\rm E}\left[\mu^2(4)\right] \left[{\bf \Lambda}^2  + {\boldsymbol \lambda \boldsymbol \lambda}^T\right],
\end{eqnarray*}
where the weighting matrix ${\bf A}(3) = {\bf F}(2) {\bf A}(2)$. Now, from the third and fourth iterative updates, we generalize the recursion for the $i^{th}$ update as
\begin{eqnarray}
\nonumber {\rm E} \left[ \left\| {\bf \bar w}(i) \right\|^2_{{\boldsymbol \sigma}} \right] \hspace{-0.6cm} &&= \left\| {\bf \bar w}_o \right\|^2_{{\bf F}(i-1){\bf A}(i-1) {\boldsymbol \sigma}} + \sigma^2_v {\rm E} \left[ \mu^2(i-1) \right] {\boldsymbol \lambda}^T {\boldsymbol \sigma} \label{mean_sq_i} \\
&&+ \sigma^2_v {\boldsymbol \lambda}^T \left\{ {\sum\limits_{k = 0}^{i-2} {{\rm E}\left[ {\mu ^2 \left( k \right)} \right]\prod\limits_{m = i-1}^{k + 1} {{\bf{F}}\left( m \right)} } } \right\} {\boldsymbol \sigma} \\
\nonumber {\bf F}(i) \hspace{-0.6cm} &&= {\rm {\bf I}}_M - 2{\rm E}\left[\mu(i)\right] {\bf \Lambda} + {\rm E}\left[\mu^2(i)\right] \left[{\bf \Lambda}^2  + {\boldsymbol \lambda \boldsymbol \lambda}^T\right].\\ \label{F_i}
\end{eqnarray}
Similarly, the recursion for the $(i+1)^{th}$ update is given by
\begin{eqnarray}
{\rm E} \left[ \left\| {\bf \bar w}(i+1) \right\|^2_{{\boldsymbol \sigma}} \right] \hspace{-0.6cm} &&= \left\| {\bf \bar w}_o \right\|^2_{{\bf F}(i){\bf A}(i) {\boldsymbol \sigma}} + \sigma^2_v {\rm E} \left[ \mu^2(i) \right] {\boldsymbol \lambda}^T {\boldsymbol \sigma} \label{mean_sq_i1} \\
\nonumber &&+ \sigma^2_v {\boldsymbol \lambda}^T \left\{ {\sum\limits_{k = 0}^{i-1} {{\rm E}\left[ {\mu ^2 \left( k \right)} \right]\prod\limits_{m = i}^{k + 1} {{\bf{F}}\left( m \right)} } } \right\} {\boldsymbol \sigma} \\
\nonumber {\bf F}(i+1) \hspace{-0.6cm} &&= {\rm {\bf I}}_M - 2{\rm E}\left[\mu(i+1)\right] {\bf \Lambda} \\
&&+ {\rm E}\left[\mu^2(i+1)\right] \left[{\bf \Lambda}^2  + {\boldsymbol \lambda \boldsymbol \lambda}^T\right]. \label{F_i1}
\end{eqnarray}
Subtracting (\ref{mean_sq_i}) from (\ref{mean_sq_i1}) and simplifying the terms gives the final recursive update equation
\begin{eqnarray}\label{mean_sq_final}
\nonumber {\rm E} \left[ \left\| {\bf \bar w}(i+1) \right\|^2_{{\boldsymbol \sigma}} \right] \hspace{-0.6cm} &&= {\rm E} \left[ \left\| {\bf \bar w}(i) \right\|^2_{{\boldsymbol \sigma}} \right] \\
\nonumber &&+ \left\| {\bf \bar w}_o \right\|^2_{{\bf F} (i) {\bf A}(i) {\boldsymbol \sigma}} + \sigma^2_v {\rm E} \left[ \mu^2(i) \right] {\boldsymbol \lambda}^T {\boldsymbol \sigma} \\
&&+ \sigma^2_v {\boldsymbol \lambda}^T \left\{ {\bf F}(i) - {\rm {\bf I}}_M \right\} {\bf B} (i) {\boldsymbol \sigma},
\end{eqnarray}
where
\begin{equation}\label{B_update1}
{\bf B}(i) = \left\{ {\rm E} \left[ \mu^2(i-1) \right]{\rm {\bf I}}_M + {\sum\limits_{k = 0}^{i-2} {{\rm E}\left[ {\mu ^2 \left( k \right)} \right]\prod\limits_{m = i-1}^{k + 1} {{\bf{F}}\left( m \right)} } } \right\}.
\end{equation}

The final set of iterative equations for the mean-square learning curve are given by (\ref{mean_sq_final}), (\ref{F_i}) and
\begin{eqnarray}
{\bf A} (i+1) \hspace{-0.6cm} &&= {\bf F} (i) {\bf A} (i) \label{A_update} \\
{\bf B} (i+1) \hspace{-0.6cm} &&= {\rm E} \left[ \mu^2 (i) \right] {\rm {\bf I}}_M + {\bf F} (i) {\bf B} (i). \label{B_update}
\end{eqnarray}
Taking the weighting matrix ${\bf \Sigma} = {\rm {\bf I}}_M$ results in the mean-square-deviation (MSD) while taking the weighting matrix ${\bf \Sigma} = {\bf \Lambda}$ gives the EMSE.

It should be noted here that unlike the analysis given in \cite{Sayedbook} for the LMS algorithm, the weighting matrix ${\bf F} (i)$ is not constant. As a result, the Cayley-Hamilton theorem is not applicable. In this context, (\ref{F_i}) and (\ref{mean_sq_final})-(\ref{B_update}) are very significant contributions of this work.

\subsection{Steady-State Analysis}
\label{SSAn}

At steady-state, the recursions (\ref{mean_sq_3}) and (\ref{F_recur}) become
\begin{eqnarray}
{\rm E}\left[ \left\| {\bf \bar w}_{ss} \right\|^2 _{{\boldsymbol \sigma}} \right] = {\rm E}\left[ \left\| {\bf \bar w}_{ss} \right\|^2 _{{\bf F}_{ss} {\boldsymbol \sigma}} \right] + \sigma^2_v \mu^2_{ss} {\boldsymbol \lambda}^T {\boldsymbol \sigma} \label{mean_sq_ss} \\
{\bf F}_{ss} = {\rm {\bf I}}_M - 2 \mu_{ss} {\bf \Lambda} + \mu^2_{ss} \left[{\bf \Lambda}^2  + {\boldsymbol \lambda \boldsymbol \lambda}^T\right], \label{F_ss}
\end{eqnarray}
where the subscript $ss$ denotes steady-state. Simplifying (\ref{mean_sq_ss}) further gives
\begin{equation}
{\rm E}\left[ \left\| {\bf \bar w}_{ss} \right\|^2 _{{\boldsymbol \sigma}} \right] = \sigma^2_v \mu^2_{ss} {\boldsymbol \lambda}^T \left[ {\rm {\bf I}}_M - {\bf F}_{ss} \right]^{-1} {\boldsymbol \sigma}, \label{mean_sq_ss1}
\end{equation}
which defines the steady-state MSD if ${\bf \Sigma} = {\rm {\bf I}}_M$ and steady-state EMSE if ${\bf \Sigma} = {\bf \Lambda}$.

\subsection{Steady-State Step-Size Analysis}
\label{step_an}

The analysis presented in the above section has been generic for any VSS algorithm. In this section, 5 different VSS algorithms are chosen to present the steady-state analysis for the step-size. These steady-state step-size values are then directly inserted into (\ref{mean_sq_ss1}) and (\ref{F_ss}). The 5 different VSS algorithms and their step-size update equations are given in Table \ref{Table1}. The first algorithm, denoted {\textbf KJ} is the work of Kwong and Johnston \cite{kwong1992}. The second algorithm, denoted by {\textbf AM}, also refers to the authors Aboulnasr and Mayyas \cite{mayyas1997}. The NC algorithm refers to the noise-constrained LMS algorithm \cite{wei2001}. The VSQ algorithm is the variable step-size quotient LMS algorithm, based on the quotient form \cite{Zhao2009}. The last algorithm, denoted by Sp, refers to the Sparse VSSLMS algorithm of \cite{Binsaeed2013}.
\begin{table}[h]
\begin{center}
\begin{tabular}{|c|l|} \hline
\textbf{Algorithm} & \textbf{Step-size update equation}\\
\hline
\textbf{KJ} \cite{kwong1992} & $\mu \left(i+1\right)=\alpha_{kj}\mu\left(i\right)+\gamma_{kj} e^{2}\left(i\right)$ \\
\hline
\textbf{AM} \cite{mayyas1997} & $p\left( i \right) = \beta_{am} p\left( {i - 1} \right) + \left( {1 - \beta_{am} } \right)e\left( i \right)e\left( {i - 1} \right)$ \\
 & $\mu \left( {i + 1} \right) = \alpha_{am} \mu \left( i \right) + \gamma_{am} p^2 \left( i \right)$ \\
\hline
\textbf{NC} \cite{wei2001} & $\mu \left( {i + 1} \right) = \mu_0 \left( 1 + \gamma_{nc} \theta_{nc} (i+1) \right)$ \\
& $\theta_{nc} (i+1) = (1 - \alpha_{nc}) \theta_{nc} (i) + \frac{\alpha_{nc}}{2} \left( e^2(i) - \sigma^2_v \right)$ \\
\hline
& $A_q(i) = a_qA_q(i-1) + e^2(i)$ \\
\textbf{VSQ} \cite{Zhao2009} & $B_q(i) = b_qB_q(i-1) + e^2(i)$ \\
& $\theta_q (i) = \frac{A_q(i)}{B_q(i)}$ \\
& $\mu \left(i+1\right)=\alpha_q\mu\left(i\right)+\gamma_q \theta_q \left(i\right)$ \\
\hline
\textbf{Sp} \cite{Binsaeed2013} & $\mu \left(i+1\right)=\alpha_{sp}\mu\left(i\right)+\gamma_{sp} \left|e\left(i\right)\right|$ \\
\hline
\end{tabular}
\end{center}
\caption{Step-size update equations for the VSSLMS algorithms.}
\label{Table1}
\end{table}

Applying the expectation operator to the step-size update equations and simplifying gives the equations presented in Table \ref{Table1E}.
\begin{table}[h]
\begin{center}
\begin{tabular}{|c|l|} \hline
\textbf{Algorithm} & \textbf{Expectation of update equation}\\
\hline
\textbf{KJ} \cite{kwong1992} & ${\rm E} \left[ \mu \left(i+1\right) \right] =\alpha_{kj} {\rm E} \left[ \mu\left(i\right) \right] + \gamma_{kj} \left[ \zeta (i) + \sigma^2_v \right].$ \\
\hline
\textbf{AM} \cite{mayyas1997} & ${\rm E} \left[ p^2\left( i \right) \right] = \beta^2_{am} {\rm E} \left[ p\left( {i - 1} \right) \right] + \left( {1 - \beta_{am} } \right)^2 \left( \zeta (i) + \sigma^2_v \right)$\\
 &\hspace{1.5cm}$ .\left( \zeta (i-1) + \sigma^2_v \right)$ \\
 & ${\rm E} \left[ \mu \left( {i + 1} \right) \right] = \alpha_{am} {\rm E} \left[ \mu \left( i \right) \right] + \gamma_{am} {\rm E} \left[ p^2 \left( i \right) \right]$ \\
\hline
\textbf{NC} \cite{wei2001} & ${\rm E} \left[ \theta_{nc} (i+1) \right] = (1 - \alpha_{nc}) {\rm E} \left[ \theta_{nc} (i) \right] + \alpha_{nc} \zeta(i) /2$ \\
& ${\rm E} \left[ \mu \left( {i + 1} \right) \right] = \mu_0 \left( 1 + \gamma_{nc} {\rm E} \left[ \theta_{nc} (i+1) \right) \right] $ \\
\hline
\textbf{VSQ} \cite{Zhao2009} & ${\rm E} \left[ \mu \left(i+1\right) \right] = \alpha_q {\rm E} \left[ \mu\left(i\right) \right] + \gamma_q \frac{a_q {\rm E} \left[ A_q(i-1) \right] + \zeta (i) + \sigma^2_v}{b_q {\rm E} \left[ B_q(i-1) \right] + \zeta (i) + \sigma^2_v} $ \\
\hline
\textbf{Sp} \cite{Binsaeed2013} & ${\rm E} \left[ \mu \left(i+1\right) \right] = \alpha_{sp} {\rm E} \left[ \mu\left(i\right) \right] + \gamma_{sp} \sqrt {2 \sigma_v^2/ \pi} $ \\
\hline
\end{tabular}
\end{center}
\caption{Expected values for the update equations from Table \ref{Table1}.}
\label{Table1E}
\end{table}

At steady-state, the expected step-size ${\rm E} \left[ \mu \left(i\right) \right]$ is replaced by $\mu_{ss}$. The approximate steady-state step-size equations are given in Table \ref{Table1S}. The steady-state EMSE (denoted by $\zeta$ in the tables) value is assumed to be small enough to be ignored.
\begin{table}[h]
\begin{center}
\begin{tabular}{|c|l|} \hline
\textbf{Algorithm} & \textbf{Steady-state step-size value}\\
\hline
\textbf{KJ} \cite{kwong1992} & $\mu_{ss} \approx \frac{\gamma_{kj}}{1-\alpha_{kj}} \sigma^2_v.$ \\
\hline
\textbf{AM} \cite{mayyas1997} & $\mu_{ss} \approx \frac{\gamma_{am} (1 - \beta_{am})}{1-\alpha_{am}} \sigma^2_v.$ \\
\hline
\textbf{NC} \cite{wei2001} & $\mu_{ss} \approx \mu_0.$ \\
\hline
\textbf{VSQ} \cite{Zhao2009} & $\mu_{ss} \approx \frac{\gamma_{q} (1 - b_q)}{1-\alpha_{q} (1 - a_q)}.$ \\
\hline
\textbf{Sp} \cite{Binsaeed2013} & $\mu_{ss} \approx \frac{\gamma_{sp} }{1-\alpha_{sp}} \sqrt {2 \sigma_v^2/ \pi}.$ \\
\hline
\end{tabular}
\end{center}
\caption{Steady-state step-size values for equations from Table \ref{Table1}.}
\label{Table1S}
\end{table}

\section{Results and Discussion}
\label{results}

In this section, the analysis presented above will be tested upon the 5 VSS algorithms listed in Table \ref{Table1}. These algorithms are used in 2 different experiments to test the validity of the analysis. In the first experiment, MSD is plotted using (\ref{mean_sq_final}) and compared with simulation results. The second experiment compares the steady-state simulation results with the theoretical results obtained using (\ref{mean_sq_ss1}).

\begin{figure}[h]
\centering{\includegraphics[height=70mm]{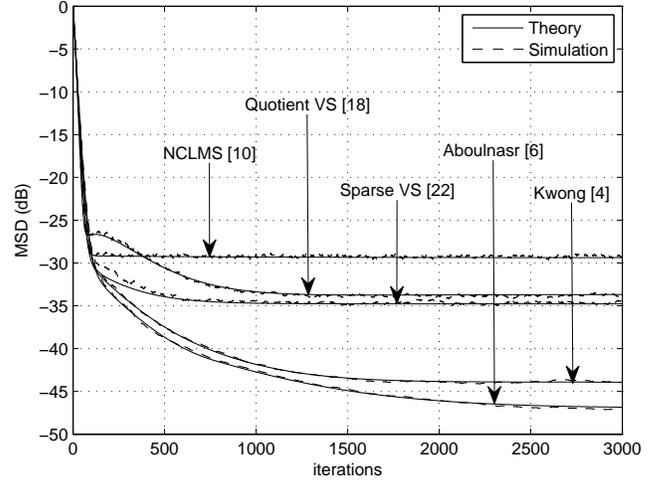}}
\caption{Theory (\ref{mean_sq_final}) v simulation MSD comparison for 5 different VSS algorithms .}\label{th_v_sim_all}
\end{figure}

For the first experiment, the length of the unknown vector is $M=4$. The input regressor vector is a realization of a zero-mean Gaussian random variable with unit variance. The signal-to-noise ratio (SNR) is chosen to be 20 dB. The step-size control parameters chosen for this experiment are given in Table \ref{Table2}. The results are shown in Fig. \ref{th_v_sim_all}. As can be seen from the figure, there is a close match between simulation and analytical results for all the algorithms.
\begin{table}[h]
\begin{center}
\begin{tabular}{|c|l|} \hline
\textbf{Algorithm} & \textbf{Parameters}\\
\hline
\textbf{KJ} \cite{kwong1992} & $\alpha_{kj} = 0.995, \gamma_{kj} = 1e-3$ \\
\hline
\textbf{AM} \cite{mayyas1997} & $\beta_{am} = 0.9, \alpha_{am} = 0.995, \gamma_{am} = 1e-3$ \\
\hline
\textbf{NC} \cite{wei2001} & $\gamma_{nc} = 10, \alpha_{nc} = 1e-3$ \\
\hline
\textbf{VSQ} \cite{Zhao2009} & $a = 0.99, b_q = 1e-3, \alpha_q = 0.995, \gamma_q = 1e-3$ \\
\hline
\textbf{Sp} \cite{Binsaeed2013} & $\alpha_{sp} = 0.995, \gamma_{sp} = 1e-3$ \\
\hline
\end{tabular}
\end{center}
\caption{Step-size control parameters for the VSSLMS algorithms.}
\label{Table2}
\end{table}

The second experiment compares simulation results for the VSS algorithms with theoretical steady-state MSD results obtained using (\ref{mean_sq_ss1}). The step-size control parameters are chosen the same as the previous experiment. The results for this experiment are given in Table \ref{Table3}. It can be seen that there is an excellent match between theory and simulation results.
\begin{table}[h]
\begin{center}
\begin{tabular}
{|c|c|c|} \hline
\textbf{Algorithm} & \textbf{MSD (dB)} & \textbf{MSD (dB)} \\
 & \textbf{equation (\ref{mean_sq_ss1})} & \textbf{simulation}\\
\hline
\textbf{KJ} \cite{kwong1992} & $-43.96$ & $-43.94$ \\
\hline
\textbf{AM} \cite{mayyas1997} & $-46.98$ & $-46.98$ \\
\hline
\textbf{NC} \cite{wei2001} & $-29.42$ & $-29.26$ \\
\hline
\textbf{VSQ} \cite{Zhao2009} & $-33.76$ & $-33.77$ \\
\hline
\textbf{Sp} \cite{Binsaeed2013} & $-34.78$ & $-34.72$ \\
\hline
\end{tabular}
\end{center}
\caption{Theory v simulation comparison for steady-state MSD for different VSSLMS algorithms.}
\label{Table3}
\end{table}


\section{Conclusion}
\label{conc}

This work presents a unified approach for the theoretical analysis of LMS-based VSS algorithms. The iterative recursions presented here differentiate this work from previous analyses in that this set of equations provides a generic treatment of the analysis for this class of algorithms for the first time. Simulation results confirm the generic behavior of the presented work, for both the transient state as well as steady-state.

\end{document}